\documentclass[12pt,a4paper]{article}
\usepackage[dvips]{graphicx}

\begin{document}

\begin{center}
\large \textbf{High domain wall velocities induced by current in
ultrathin Pt/Co/AlOx wires with perpendicular magnetic anisotropy}

\vspace{1cm}

\normalsize

\centering T.A. Moore, I.M. Miron, G. Gaudin, G. Serret, S. Auffret,
B. Rodmacq, \\A. Schuhl

\centering \emph{SPINTEC, URA 2512, CEA/CNRS, CEA/Grenoble,\\ 38054
Grenoble Cedex 9, France}

\vspace{0.5cm}

\centering S. Pizzini, J. Vogel

\centering \emph{Institut N\'eel, CNRS and UJF, B.P. 166, 38042
Grenoble Cedex 9, France}

\vspace{0.5cm}

\centering M. Bonfim

\centering \emph{Departamento de Engenharia El\'etrica, Universidade
Federal do Paran\'a, Curitiba, Paran\'a, Brazil}

\vspace{0.5cm}

\end{center}

\textbf{Abstract}

Current-induced domain wall (DW) displacements in an array of
ultrathin Pt/Co/AlOx wires with perpendicular magnetic anisotropy
have been directly observed by wide field Kerr microscopy.  DWs in
all wires in the array were driven simultaneously and their
displacement on the micrometer-scale was controlled by the current
pulse amplitude and duration. At the lower current densities where
DW displacements were observed ($j \leq 1.5 \times 10^{12}$
A/m$^{2}$), the DW motion obeys a creep law. At higher current
density ($j = 1.8 \times 10^{12}$ A/m$^{2}$), zero-field average DW
velocities up to $130 \pm 10$ m/s were recorded.

\newpage

Magnetic domain wall (DW) propagation by spin-polarized current,
predicted by Berger \cite{Berger84}, has attracted huge attention in
the last few years (see \cite{Beach08, Tserkovnyak08} and references
therein) due to unsolved questions about the underlying wall
propagation mechanism and the possibility of applications in
spintronic devices \cite{Parkin08}. Up to now current-induced domain
wall motion has been mainly studied in flat nanoscale strips or
``wires'' of NiFe (Permalloy) \cite{Vernier04, Yamaguchi04,
Hayashi07, Meier07} where the magnetization is oriented along the
length of the wire. However, the potential disadvantages of using
Permalloy wires in devices include the difficulty of achieving fast,
controllable DW motion, and the large drive currents required.
Although a DW velocity due to current of 110 m/s has been reported
\cite{Hayashi07, Meier07}, it has also been demonstrated that the DW
motion is stochastic due to thermal effects and local pinning
\cite{Klaui05b}, and that the DW may undergo spin structure
transformations \cite{Heyne08}, leading to complex dynamics.
Micromagnetic simulations have shown that in metallic wires with
perpendicular magnetic anisotropy, these problems may be overcome
\cite{Jung08, Fukami08}. However, in very few experiments on wires
with perpendicular anisotropy have current-induced DW displacements
been observed, probably due to the strong intrinsic pinning. In
experiments that have shown current-induced DW displacements the DW
velocities are generally smaller than in Permalloy, e.g. Tanigawa et
al. \cite{Tanigawa08} obtained a DW velocity of 0.05 m/s in a CoCrPt
wire, while Koyama et al. \cite{Koyama08} reported 40 m/s in a Co/Ni
wire.

We study current-induced DW motion in Pt/Co/AlOx nanowires and
compare the results with similar measurements on Pt/Co/Pt wires.  In
these systems the magnetization in the Co layer points out of the
plane \cite{Manchon08, Metaxas07}, narrow ($\sim$10 nm) Bloch-type
DWs occur and a high spin torque efficiency is expected
\cite{Jung08, Fukami08}. Apart from the top layer (AlOx or Pt), the
two types of nanowire in our study have the same structure. The
presence of the AlOx in the Pt/Co/AlOx system, breaking the
inversion symmetry, is expected to enhance the spin torque via an
increase of the spin flip rate, and this effect was recently
evidenced for DW displacements on the order of a few nanometers
\cite{Miron08}. In this Letter, we examine the effect of this
symmetry breaking on micrometer-scale DW displacements.

\vspace{0.5cm}

The nanowires are 500 nm wide and approximately 10 $\mu$m long,
patterned from magnetron sputtered films of
Pt(3nm)/Co(0.6nm)/AlOx(2nm) or Pt(3nm)/Co(0.6nm)/Pt(3nm) on
Si/SiO$_{2}$(500nm) by electron beam lithography and Ar ion etching.
Twenty wires are arranged in parallel with a repeat distance of 2
$\mu$m and connected at each end to a micrometer-scale domain
nucleation pad. Fig.~1(a) shows part of the wire array. An Au
contact is defined by optical lithography on top of each nucleation
pad, giving a total wire array resistance $\sim100$ $\Omega$.

The magnetization of the wire array is saturated out-of-plane in an
external field of about 4 kOe, and then applying a reverse field,
DWs nucleated in the pads propagate into the wires. By precisely
controlling the field strength, DWs are positioned in the wires
(Fig.~1(a)). Subsequently the field is decreased to zero and a
number of current pulses $n$ (nominal pulse length $t = 0.8-5$ ns,
density $j$ up to $1.8 \times 10^{12}$ A/m$^{2}$) are injected into
the wires via the Au contacts.

\vspace{0.5cm}

Using wide field Kerr microscopy in differential imaging mode, the
current-induced displacement of DWs in the Pt/Co/AlOx wires was
directly observed.  Fig.~1(b) shows two Pt/Co/AlOx wires with two
DWs in each wire. The image is the difference between the domain
pattern in the initial state and in the state after the current
pulse has been applied, and thus there is contrast only in regions
where the magnetization in the wire has reversed from one direction
to the other. The initial wall positions are marked by dotted lines
and the electron flow is from right to left. The displacement of the
DWs gives black or white contrast depending on whether the
magnetization reverses from ``up'' to ``down'' or vice versa. The DW
displacement can be controlled by varying the amplitude or length of
the current pulse or the number of pulses. When the current
direction is reversed, DW motion continues to be opposite to the
direction of electron flow (i.e. also reversed) and DW displacements
are of the same magnitude. These results for Pt/Co/AlOx wires are in
contrast with measurements on Pt/Co/Pt wires, where DW displacements
were not observed. The only effect of the current in the Pt/Co/Pt
wires was, for sufficiently long pulses of high amplitude (e.g. $t =
5$ ns, $j = 1.6 \times 10^{12}$ A/m$^{2}$), to nucleate reverse
domains at random locations along the wire due to Joule heating.

Domain nucleation is also observed in the Pt/Co/AlOx wires, but only
if the current pulse is long enough. In these wires the rate of
Joule heating is not as high because an insulating AlOx layer has
replaced the Pt top layer and thus less current is required for a
given $j$. If the current pulse is short (e.g. $t = 0.8$ ns),
thermal equilibrium is unlikely to be reached and the temperature
$T$ of the wire remains well below the Curie temperature ($T_{C}
\sim$ 500 K). The evidence for this is that we do not observe domain
nucleation even at the highest current density in the Pt/Co/AlOx
wires. In the experiment we decrease the pulse length as the current
density is increased in order to ensure that $T << T_{C}$.

We capture ten images of DW displacements in the Pt/Co/AlOx wires
for each value of current density.  Fig.~2 shows probability
distributions, each built from approximately 200 DW motion events,
of the DW velocity for different values of current density.  At $j =
1.0 \times 10^{12}$ A/m$^{2}$, the lowest current density for which
DW displacements were observed, 500 pulses of 5 ns duration were
required in order to produce a DW displacement on the order of
$\Delta = 500$ nm.  This corresponds to low DW velocities of $v =
\Delta/(n t) = 0.27 \pm 0.02$ m/s on average, as shown in Fig.~2(a).
Meanwhile, for $j = 1.8 \times 10^{12}$ A/m$^{2}$, the highest
attainable current density in this experiment, only ten pulses of
0.8 ns duration were needed to cause a DW displacement exceeding 1
$\mu$m, corresponding to an average DW velocity of $68 \pm 1$ m/s
(Fig.~2(b)).

As the current density increases, the DW velocity distribution
changes from one that is skewed towards low velocities (Fig.~2(a))
to one that is more symmetric (Fig.~2(b)).  Additionally, the width
of the distribution decreases (inset of Fig.~2(a)). The large width
of the distribution at low current densities indicates that in this
regime the DW motion is predominantly stochastic; then as $j$
increases, the decreasing width implies that the DW motion becomes
more reproducible.

The mean DW velocity $v$ appears to increase exponentially as a
function of the current density (Fig.~3(a)), suggesting that in the
measured range of $j$ the DW exhibits creep motion described by
\cite{Chauve00}:

\begin{equation}
v = v_{0} \exp
\bigg[-\bigg(\frac{T_{dep}}{T}\bigg)\bigg(\frac{f_{dep}}{f}\bigg)
^{\mu} \bigg].
\end{equation}

Here, $T_{dep}$ is the depinning temperature given by $U_{C}/k_{B}$,
where $U_{C}$ is related to the height of the DW pinning energy
barrier, $f_{dep}$ is the depinning force ($\equiv j_{dep}$, the
depinning current density), $v_{0}$ is a numerical prefactor and
$\mu$ is a universal dynamic exponent equal to $1/4$ for a 1D
interface moving in a 2D weakly disordered medium \cite{Chauve00}.
To check whether our DW motion obeys the creep law, we plot ln$v$
versus $j^{-1/4}$ (Fig.~3(b)). Linear behavior is seen at the lower
current densities $j = 1.0-1.5 \times 10^{12}$ A/m$^{2}$, which
verifies not only that DW creep occurs in this regime, but also that
$\mu = 1/4$ holds true for our system.

At $j \approx 1.5 \times 10^{12}$ A/m$^{2}$ there is a deviation
from the creep motion; although the DW velocity continues to
increase with further increasing current density, it increases less
rapidly than expected by the creep law. The deviation from creep
motion cannot be explained by sample heating since in this case the
creep law would predict even higher velocities. A possible
explanation is that at $j \approx 1.5 \times 10^{12}$ A/m$^{2}$ the
DW motion starts a transition from the creep to a viscous flow
regime \cite{Metaxas07}. This would result in smaller DW velocities
than expected by the creep law in the range $j = 1.5-1.8 \times
10^{12}$ A/m$^{2}$.

Fig.~3(a) also shows the maximum DW velocities measured at each
current density. A top speed of $130 \pm 10$ m/s (average over ten
pulses) was recorded at $j = 1.8 \times 10^{12}$ A/m$^{2}$, which is
large compared to previously reported values for wires with
perpendicular anisotropy \cite{Tanigawa08, Koyama08}. These high DW
velocities are surprising, given the difficulty of displacing DWs in
the same manner in Pt/Co/Pt wires.  It is reasonable to expect that
the explanation for this stems from the presence of the Co/AlOx
interface. Recently the ratio of the non-adiabatic and adiabatic
spin torque components, $\beta$, representing the spin torque
efficiency, has been experimentally determined to be of the order of
1 in Pt/Co/AlOx \cite{Miron08}, and at least 50 times smaller than
this in Pt/Co/Pt. The reason for the difference in $\beta$ is
thought to be an increase in the spin flip rate at the Co/AlOx
interface \cite{Miron08}. As $\beta$ controls the DW motion, this
difference could explain the fact that DW displacements are only
seen in Pt/Co/AlOx wires, and that the DW velocities there are high.

\vspace{0.5cm}

In summary, we have studied DW motion in ultrathin Pt/Co/AlOx
nano-wires induced by nanosecond current pulses.  DW displacements
on the micrometer-scale were observed, and could be controlled by
varying the amplitude or length of the current pulse, or the number
of pulses.  In the current density range $j = 1.0-1.5 \times
10^{12}$ A/m$^{2}$ the DWs exhibit stochastic creep motion, whereas
at higher current densities in the range $j = 1.5-1.8 \times
10^{12}$ A/m$^{2}$ the DW motion is more reproducible and velocities
greater than 100 m/s were measured, indicating a large spin torque
efficiency in this material. This result offers a route to the
realisation of magnetoelectronic devices based on current-induced DW
motion.

\vspace{0.5cm}

The authors acknowledge support of Nanofab/CNRS Institut N\'eel.
This work was partly funded by the ANR-07-NANO-034 `Dynawall'.

\newpage

\textbf{Figure captions}

\vspace{0.5cm}

\textbf{Figure 1.} (a) Wide field Kerr microscope image of part of
the Pt/Co/ AlOx wire array, with Au contacts shown schematically.
The wires are approximately 10 $\mu$m long.  The boundaries between
dark/light contrast in the wires indicate the DW positions. (b)
Differential Kerr microscope image of current-induced DW
displacements in 500 nm-wide Pt/Co/AlOx wires, driven by 10 $\times$
1.5 ns current pulses of density $j = 1.7 \times 10^{12}$ A/m$^{2}$.

\vspace{0.5cm}

\textbf{Figure 2.} Probability distributions of current-induced DW
velocity in 500 nm-wide Pt/Co/AlOx wires, for different values of
current density (a) $1.0 \times 10^{12}$ A/m$^{2}$ and (b) $1.8
\times 10^{12}$ A/m$^{2}$. The mean DW velocity is indicated in each
figure by a dashed line: (a) 0.27 m/s, (b) 68 m/s.  The inset in (a)
shows the ratio of the standard deviation $\sigma$ of the
distribution to the mean velocity $v$ as a function of the current
density $j$.

\vspace{0.5cm}

\textbf{Figure 3.} (a) Mean and maximum DW velocity as a function of
current density for 500 nm-wide Pt/Co/AlOx wires.  The error bars
are smaller than the data points unless shown. (b) The mean DW
velocity $v$ fits to a creep law at current densities $j = 1.0-1.5
\times 10^{12}$ A/m$^{2}$ (corresponding to $j^{-1/4} =
0.9-1.0~(\times 10^{12}$~A/m$^{2}$)$^{-1/4}$).

\newpage

\textbf{Figure 1.}

\vspace{1cm}

\begin{figure}[h]
  \centering
  \includegraphics[width=0.55\textwidth]{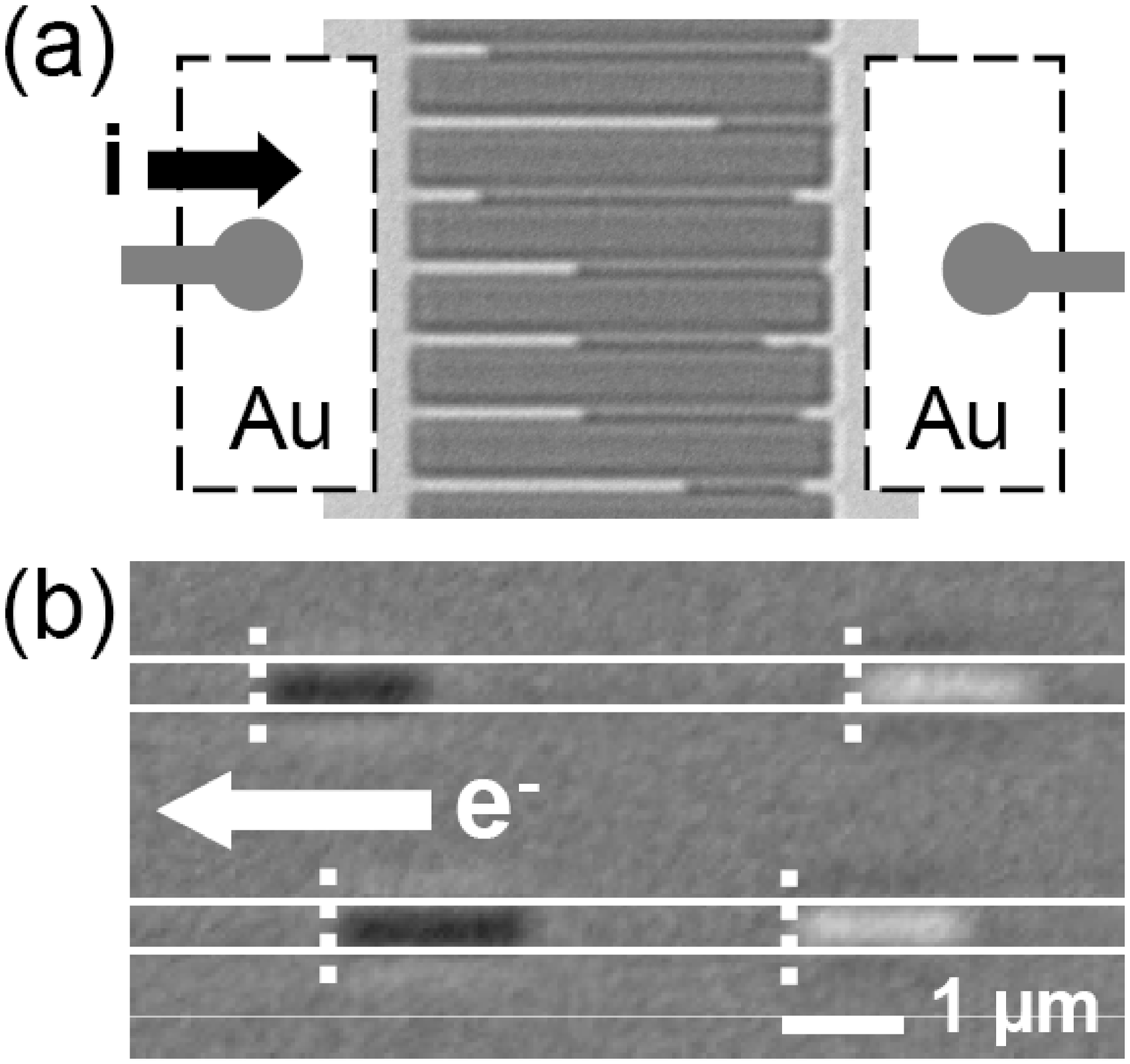}\\
\end{figure}

\newpage

\textbf{Figure 2.}

\vspace{1cm}

\begin{figure}[h]
  \centering
  \includegraphics[width=0.65\textwidth]{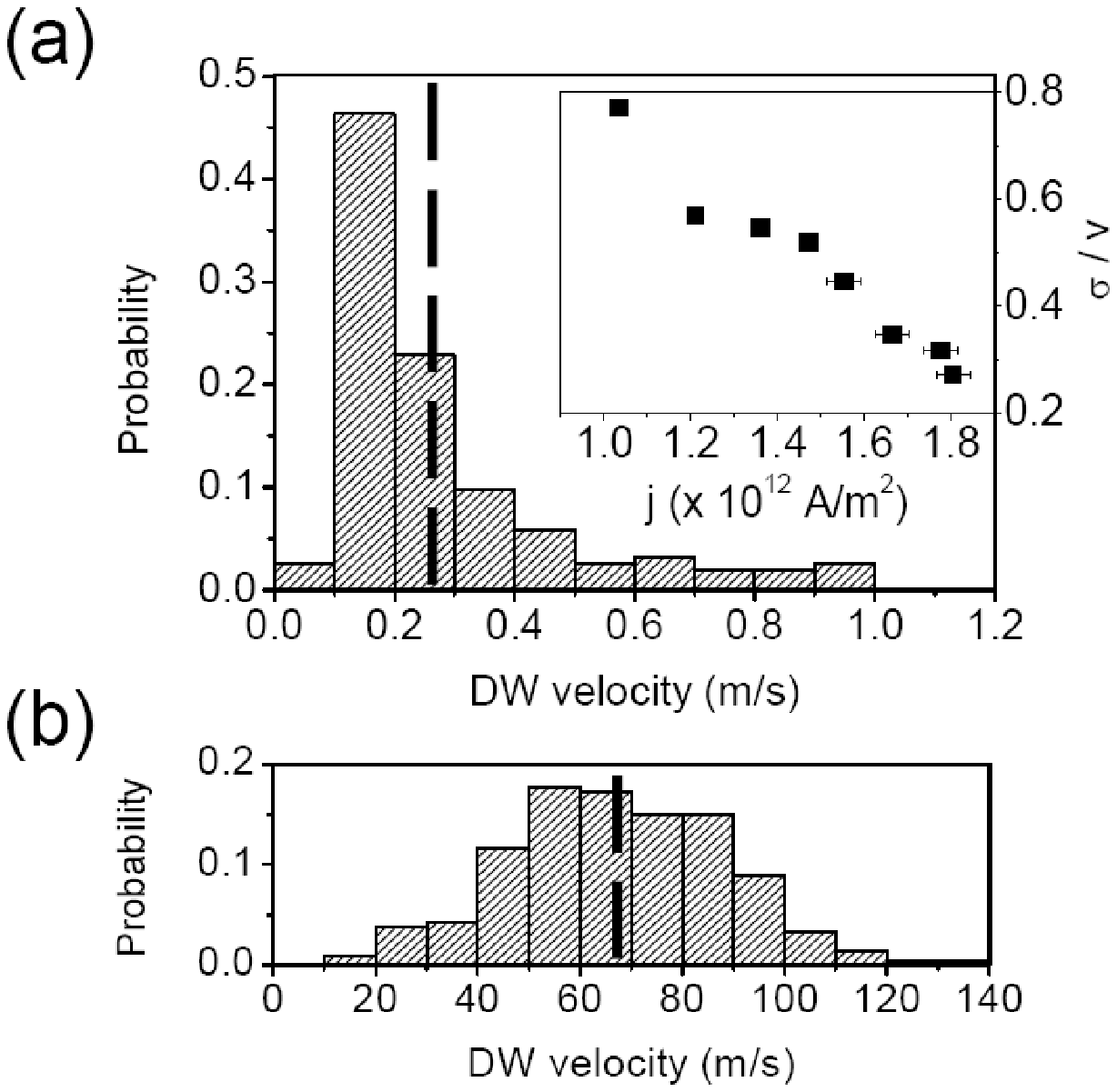}\\
\end{figure}

\newpage

\textbf{Figure 3.}

\vspace{1cm}

\begin{figure}[h]
  \centering
  \includegraphics[width=0.6\textwidth]{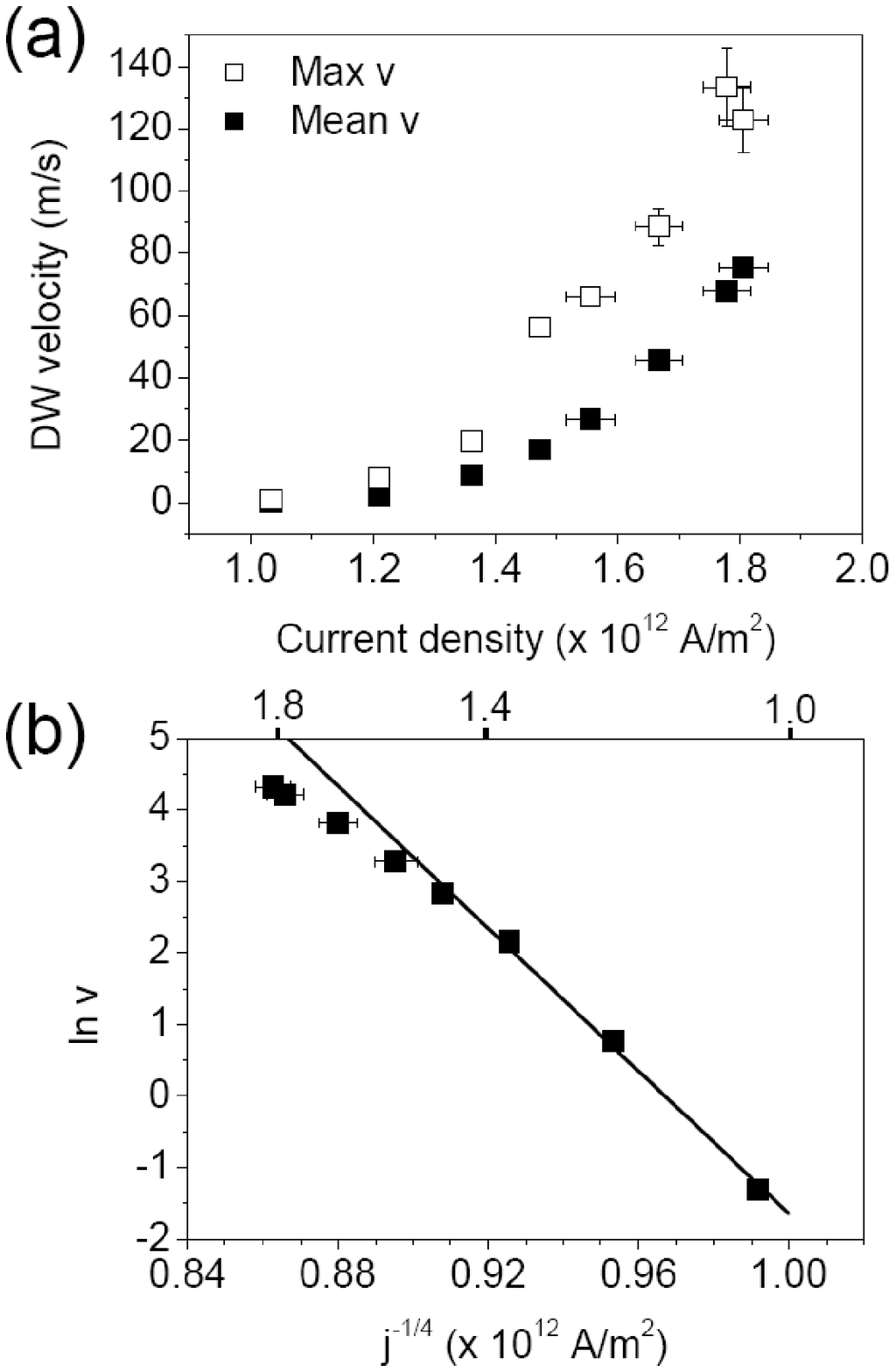}\\
\end{figure}

\end{document}